\documentclass[12pt]{article}
\usepackage[utf8]{inputenc}
\usepackage{textcomp}
\usepackage{graphicx}
\usepackage{amsmath}
\usepackage{array}
\usepackage{multirow}
\usepackage{booktabs}
\usepackage[hypcap=false]{caption}
\usepackage{xspace}
\usepackage[hidelinks]{hyperref}

\DeclareFontEncoding{LGR}{}{}
\DeclareFontSubstitution{LGR}{cmr}{m}{n}
\DeclareTextSymbol{\textgamma}{LGR}{103}
\DeclareTextSymbolDefault{\textgamma}{LGR}

\newcommand{\mybar}[1]{\smash{$\bar{\text{#1}}$}}
\newcommand{\tbar}{\mybar{t}}
\newcommand{\ttbar}{t\tbar}
\newcommand{\mygamma}{\textgamma\xspace}
\newcommand{\mymu}{\textmu\xspace}

\newcommand{\TeV}{\,TeV\xspace}
\newcommand{\sqrts}[1][13]{$\sqrt{s}=#1$\TeV}
\newcommand{\perfb}{\,fb\textsuperscript{-1}\xspace}

\begin{document}
\begin{titlepage}\pagenumbering{Alph}
\rightline{\begin{tabular}{l}
    CMS-CR-2021/289 \\ 
    December 30, 2021 
\end{tabular}}

\vfill
\begin{center}\Large 
    Top quark production in association\\ with a vector boson at the LHC
\end{center}
\vfill
\begin{center}
    \href{mailto:joscha.knolle@cern.ch}{\textsc{Joscha Knolle}} \\
    \textit{Ghent University, Ghent, Belgium}
\end{center}
\begin{center}
    \textsc{on behalf of the ATLAS \& CMS Collaborations}
\end{center}
\vfill
\begin{quotation} 
    Cross section measurements of top quark production in association with a vector boson in proton-proton collisions at \sqrts at the CERN LHC provide an important probe of the electroweak top quark couplings.
    In this contribution, recent results of the ATLAS and CMS Collaborations for the measurement of top quark pair and single top quark production in association with a photon or a Z~boson are presented.
    Both inclusive and differential cross section measurements are performed, and the results are compared to state-of-the-art predictions in the standard model of particle physics.
\end{quotation}
\vfill
\begin{quotation}\begin{center}
    PRESENTED AT
\end{center}\bigskip\begin{center}\large
    14\textsuperscript{th} International Workshop on Top Quark Physics\\
    (videoconference), 13--17 September, 2021
\end{center}\end{quotation}
\vfill
\end{titlepage}
\def\thefootnote{\fnsymbol{footnote}}
\setcounter{footnote}{0}
\pagenumbering{arabic}

\section{Introduction}

Top quark production in association with a vector boson provides an important probe of the electroweak couplings of the top quark.
As precise predictions within the standard model (SM) of particle physics become available, inclusive and differential cross section measurements can be used to test our theoretical understanding of the production mechanisms.
The experimental results are also used to constrain models of physics beyond the SM, e.g.\ through interpretations in the framework of the SM effective field theory.

Results of the ATLAS~\cite{ATLAS-Experiment} and CMS~\cite{CMS-Experiment} experiments for inclusive cross section measurements of top quark pair (\ttbar) and single top quark (t) production in association with a photon (\mygamma), a Z~boson, or a W~boson are summarized in Table~\ref{tab:summary}.
For the results, the experiments used proton-proton (pp) collision data recorded at \sqrts between 2015 and 2018 at the CERN LHC.
In this contribution, the most recent results for \ttbar\mygamma, \ttbar Z, and tZq production are presented.

\begin{center}
\renewcommand\arraystretch{1.1}
\newcommand{\mypm}{\,$\pm$\,}
\begin{tabular}{llr@{}>{\mypm}l<{\,fb}@{~}lr@{}l@{~}l}
    \toprule
    \textbf{process} & \multicolumn{4}{c}{\textbf{measurement}} & \multicolumn{3}{c}{\textbf{prediction}} \\
    \midrule
    fiducial {\ttbar\mygamma}\,+\,tW\mygamma (e\mymu) & ATLAS & 39.6 & 2.5 & \cite{ATLAS-ttG} & 38.5 & \mypm1.9\,fb & \cite{theory-ttG1,theory-ttG2} \\
    fiducial \ttbar\mygamma ($\ell$+jets) & CMS & 798 & 49 & \cite{CMS-ttGsl} & 773 & \mypm135\,fb & \\
    fiducial \ttbar\mygamma ($2\ell$) & CMS & 174.4 & 6.6 & \cite{CMS-ttGdi} & 153 & \mypm25\,fb \\
    fiducial t{\mygamma}q ({\mymu}+jets) & CMS & 115 & 30 & \cite{CMS-tGq} & 81 & \mypm4\,fb \\
    \midrule
    \multirow{2}{*}{\ttbar Z} & CMS & 950 & 80 & \cite{CMS-ttZ} & \multirow{2}{*}{859} & \multirow{2}{*}{\mypm78\,fb} & \multirow{2}{*}{\cite{theory-ttZ}} \\
    & ATLAS & 990 & 90 & \cite{ATLAS-ttZ} \\
    \multirow{2}{*}{tZq, Z\,$\to$\,$\ell\ell$} & ATLAS & 97 & 15 & \cite{ATLAS-tZq} & \multirow{2}{*}{94} & \multirow{2}{*}{\mypm3\,fb} & \multirow{2}{*}{\cite{CMS-tZq-pred}} \\
    & CMS & 87.9 & 10.0 & \cite{CMS-tZq} \\
    \midrule
    \multirow{2}{*}{\ttbar W} & CMS & 770 & 170 & \cite{CMS-ttW} & \multirow{2}{*}{722} & \multirow{2}{*}{\mypm74\,fb} & \multirow{2}{*}{\cite{theory-ttW}} \\
    & ATLAS & 870 & 190 & \cite{ATLAS-ttW} \\
    \bottomrule
\end{tabular}
\captionof{table}{%
    Summary of inclusive cross section measurements and SM predictions for associated top quark production processes.
    Dedicated SM calculations are referenced.
    Otherwise, the SM expectation has been evaluated by the experimental collaboration with an MC event generator.
}
\label{tab:summary}
\end{center}

\section{Measurements of \texorpdfstring{\ttbar\mygamma}{ttG} production}

The ATLAS Collaboration measured \mygamma-associated top quark production in e\mymu events, using 139\perfb of data~\cite{ATLAS-ttG}.
A fiducial phase space is defined at parton-level for the combined \ttbar\mygamma and tW\mygamma production as signal process.
A template fit is performed to the scalar $p_{\mathrm{T}}$ sum of the reconstructed objects (photon, leptons, jets).
The inclusive cross section is measured with a precision of 6.3\%, and found in agreement with the dedicated SM prediction from Ref.~\cite{theory-ttG1, theory-ttG2}.

The CMS Collaboration measured \ttbar\mygamma production in $\ell$+jets ($2\ell$) events, using 137\perfb (138\perfb) of data~\cite{CMS-ttGsl, CMS-ttGdi}.
A fiducial phase space for both measurements is defined at particle-level.
For the $\ell$+jets analysis~\cite{CMS-ttGsl}, a template fit is performed to the invariant mass distribution of the three selected jets, split by lepton flavour and jet multiplicity, and background contributions are constrained from control regions.
The inclusive cross section is measured with a precision of 6.1\%.
For the $2\ell$ analysis~\cite{CMS-ttGdi}, a template fit is performed to the photon $p_{\mathrm{T}}$ distribution, split by lepton flavour.
The inclusive cross section is measured with a precision of 3.8\%.
Both results are found in agreement with the SM expectation as evaluated with an MC event generator.

All three results also present differential cross section measurements, and apply unfolding procedures to correct for detector resolution effects and to extrapolate the distributions to the fiducial phase space at parton- (ATLAS) or particle-level (CMS).
Two examples are shown in Figure~\ref{fig:ttG}, and are compared to different SM predictions.

\begin{center}
\raisebox{4pt}{\includegraphics[width=0.445\textwidth]{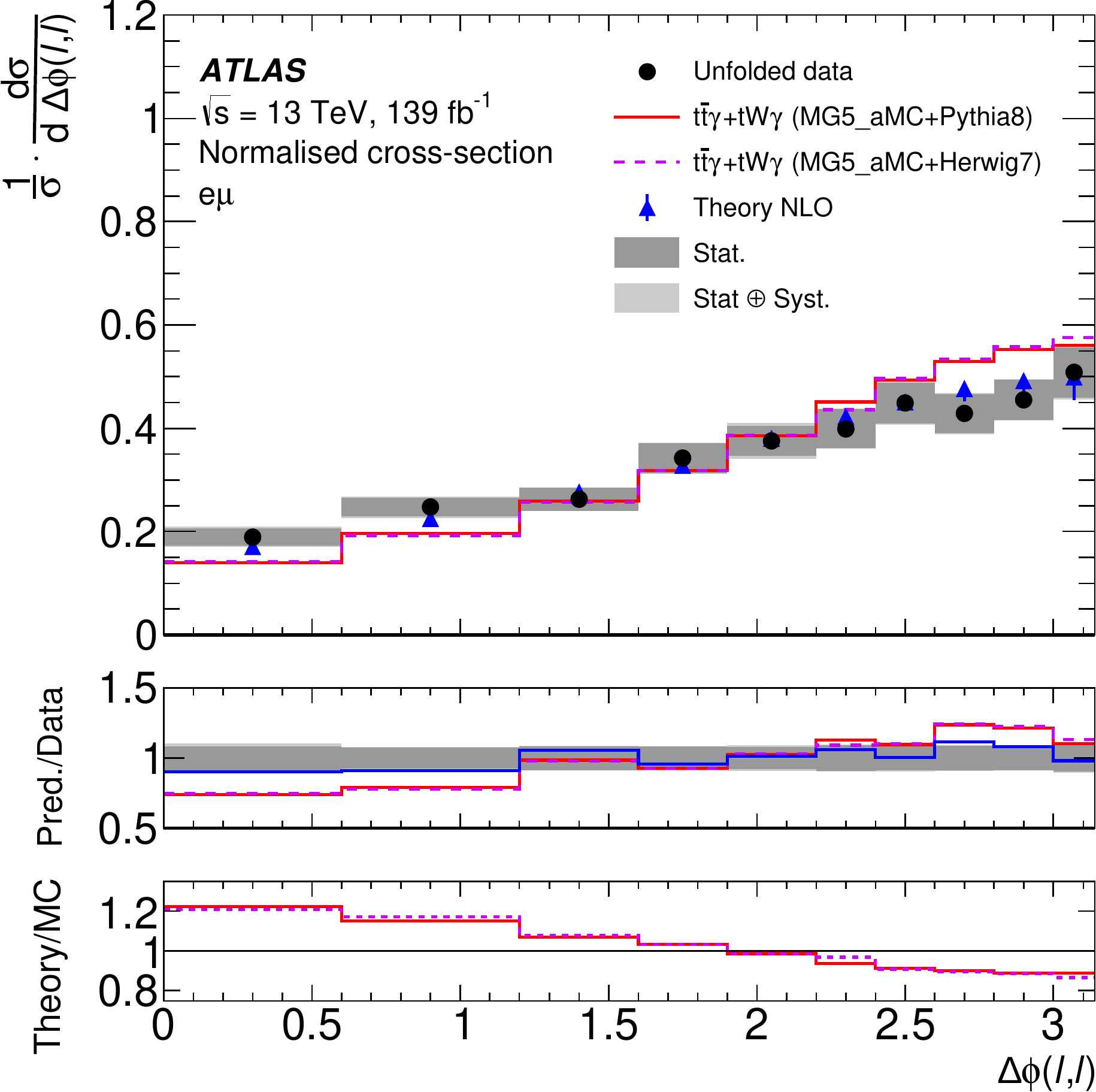}}%
\hfill
\includegraphics[width=0.5\textwidth]{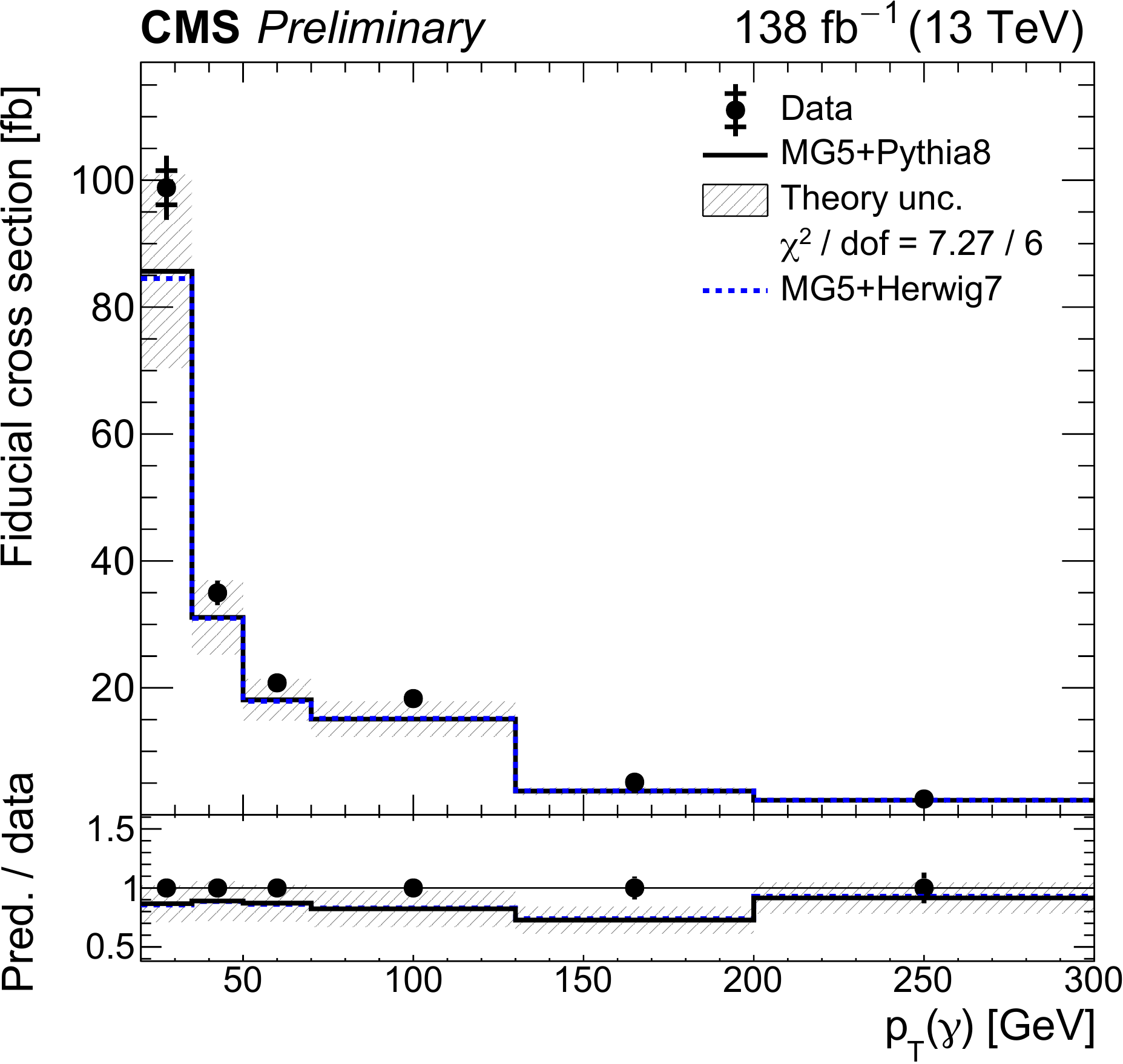}%
\captionof{figure}{%
    (Left) Normalized differential {\ttbar\mygamma}+tW\mygamma production cross section as a function of the azimuthal angle between the two leptons~\cite{ATLAS-ttG}.
    (Right) Absolute differential \ttbar\mygamma production cross section as a function of the photon $p_{\mathrm{T}}$~\cite{CMS-ttGdi}.
}
\label{fig:ttG}
\end{center}

\section{Measurements of \texorpdfstring{\ttbar Z}{ttZ} production}

The \ttbar Z production process was measured in $3\ell$ and $4\ell$ events by the CMS Collaboration using 77.5\perfb of data~\cite{CMS-ttZ}, and by the ATLAS Collaboration using 139\perfb of data~\cite{ATLAS-ttZ}.
In both measurements, a template fit is performed to the distributions of lepton, jet, and b-jet multiplicity.
The inclusive cross section is measured by the CMS Collaboration with a precision of 8.2\%, and by the ATLAS Collaboration with a precision of 9.5\%.
Both results are in agreement with each other, and also in agreement with the dedicated SM prediction from Ref.~\cite{theory-ttZ}.

In both results, differential cross section measurements are presented as well.
For the CMS result, two distributions are measured in $3\ell$ events, and an unfolding procedure is applied to correct for detector resolution effects and to extrapolate to parton-level.
For the ATLAS result, several distributions are measured for different channels, and are extrapolated to parton- or particle-level.
Two examples are shown in Figure~\ref{fig:ttZ}, and are compared to different SM predictions.

\begin{center}
\includegraphics[width=0.468\textwidth]{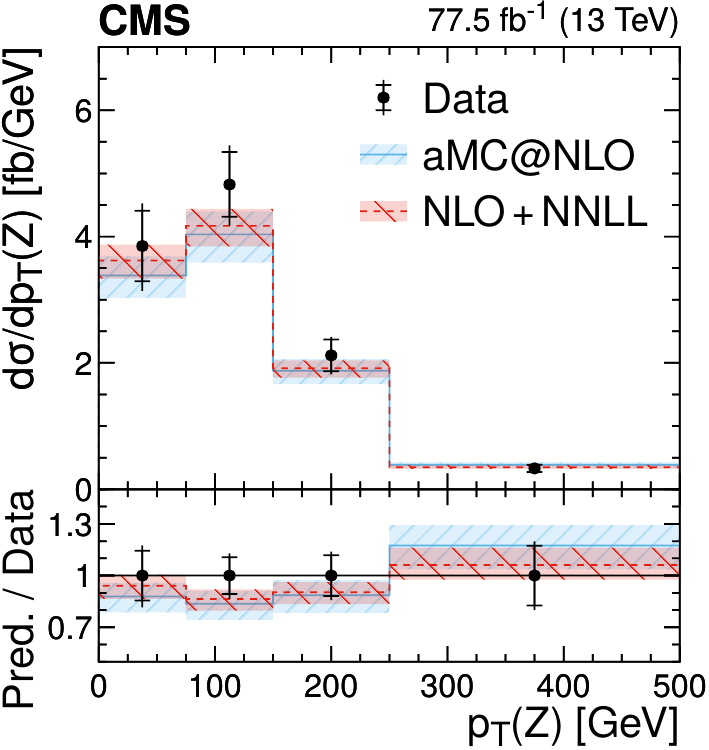}%
\hfill
\raisebox{8pt}{\includegraphics[width=0.485\textwidth]{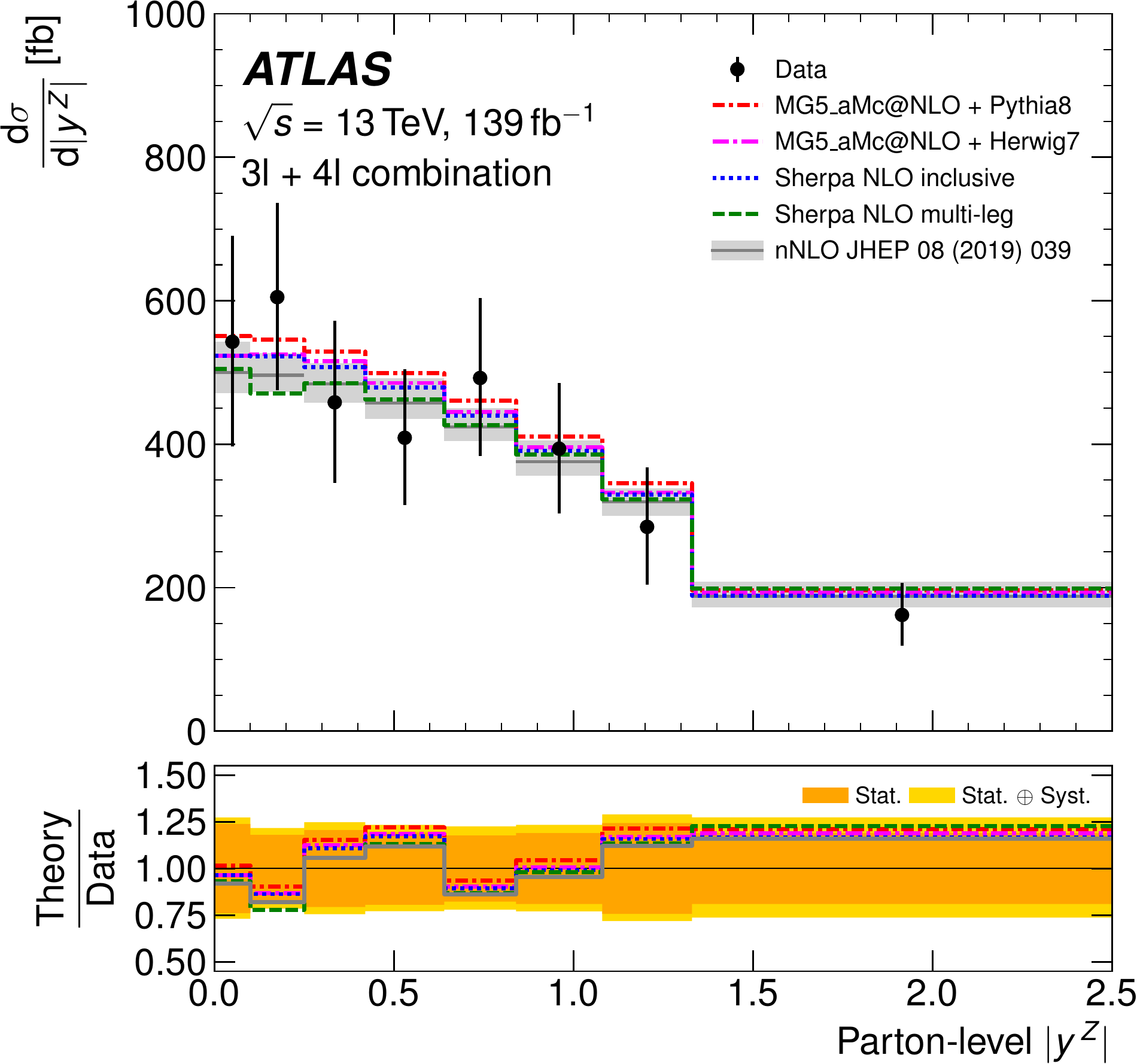}}%
\captionof{figure}{%
    (Left) Absolute differential \ttbar Z production cross section as a function of the Z~boson $p_{\mathrm{T}}$~\cite{CMS-ttZ}.
    (Right) Absolute differential \ttbar Z production cross section as a function of the Z~boson rapidity~\cite{ATLAS-ttZ}.
}
\label{fig:ttZ}
\end{center}

\section{Measurements of tZq production}

The tZq production in the $3\ell$ channel was measured by the ATLAS Collaboration in events with 2--3 jets using 139\perfb of data~\cite{ATLAS-tZq}, and by the CMS Collaboration in events with at least two jets using 138\perfb of data~\cite{CMS-tZq}.
In both measurements, a machine-learning discriminant is trained to distinguish between the tZq signal and background processes (especially \ttbar Z and diboson production).
Template fits are performed to the discriminant distributions, split by jet and b-jet multiplicities, and background contributions are constrained from control regions.
The inclusive cross section is measured by the ATLAS Collaboration with a precision of 14\%, and by the CMS Collaboration with a precision of 11\%.
Both results are in agreement with each other, and are also in agreement with the SM expectation evaluated in Ref.~\cite{CMS-tZq-pred}.

The CMS Collaboration also performed a differential cross section measurement, using a multiclass machine-learning discriminant to distinguish simultaneously between the tZq signal and the different classes of background processes.
A likelihood-based unfolding procedure is applied to correct for detector resolution effects, to extrapolate to parton- or particle-level, and to also constrain background contributions from control regions.
Two examples are shown in Figure~\ref{fig:tZq}, and are compared to different SM predictions.

\begin{center}
\includegraphics[width=0.47\textwidth]{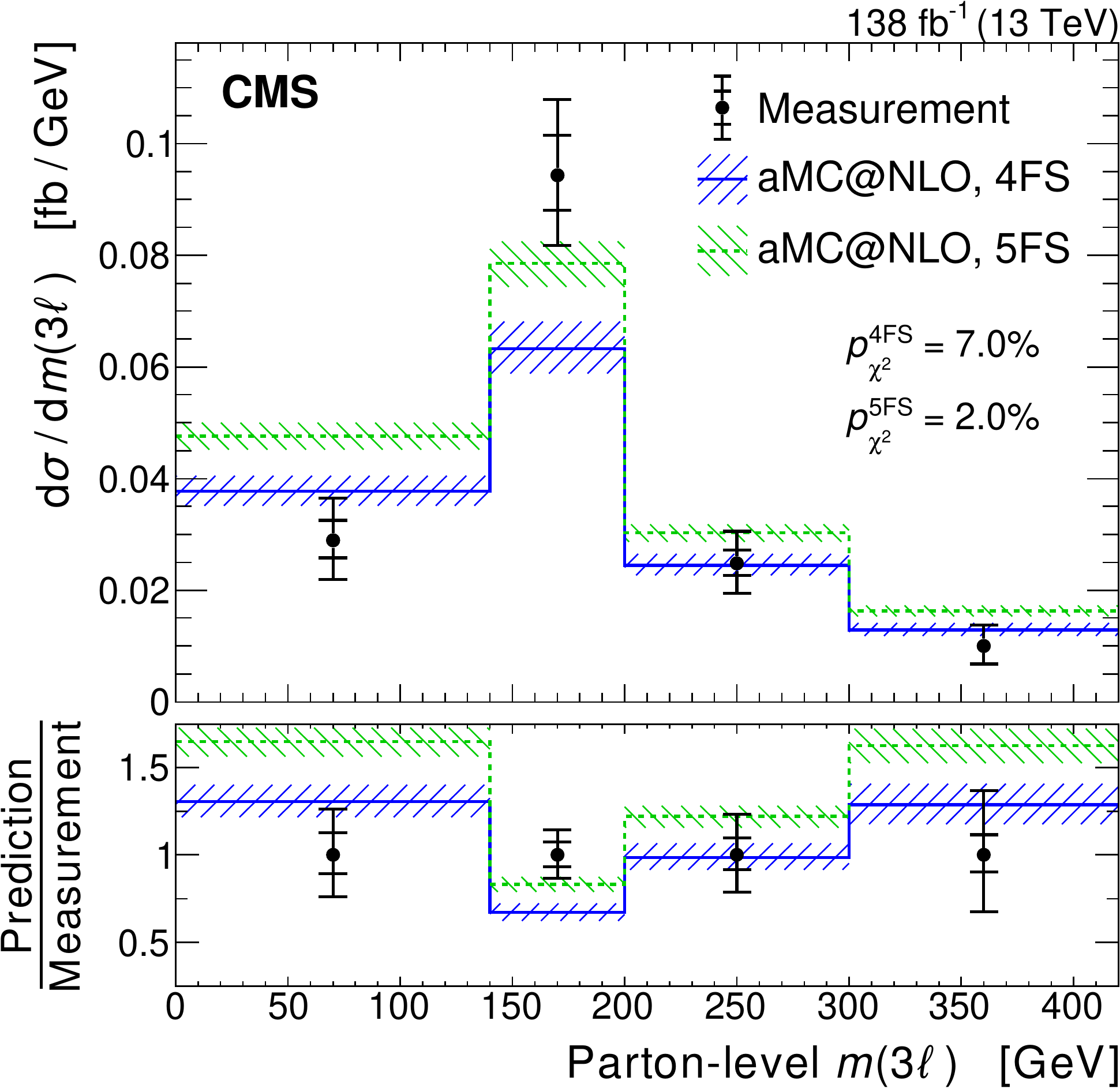}%
\hfill
\includegraphics[width=0.47\textwidth]{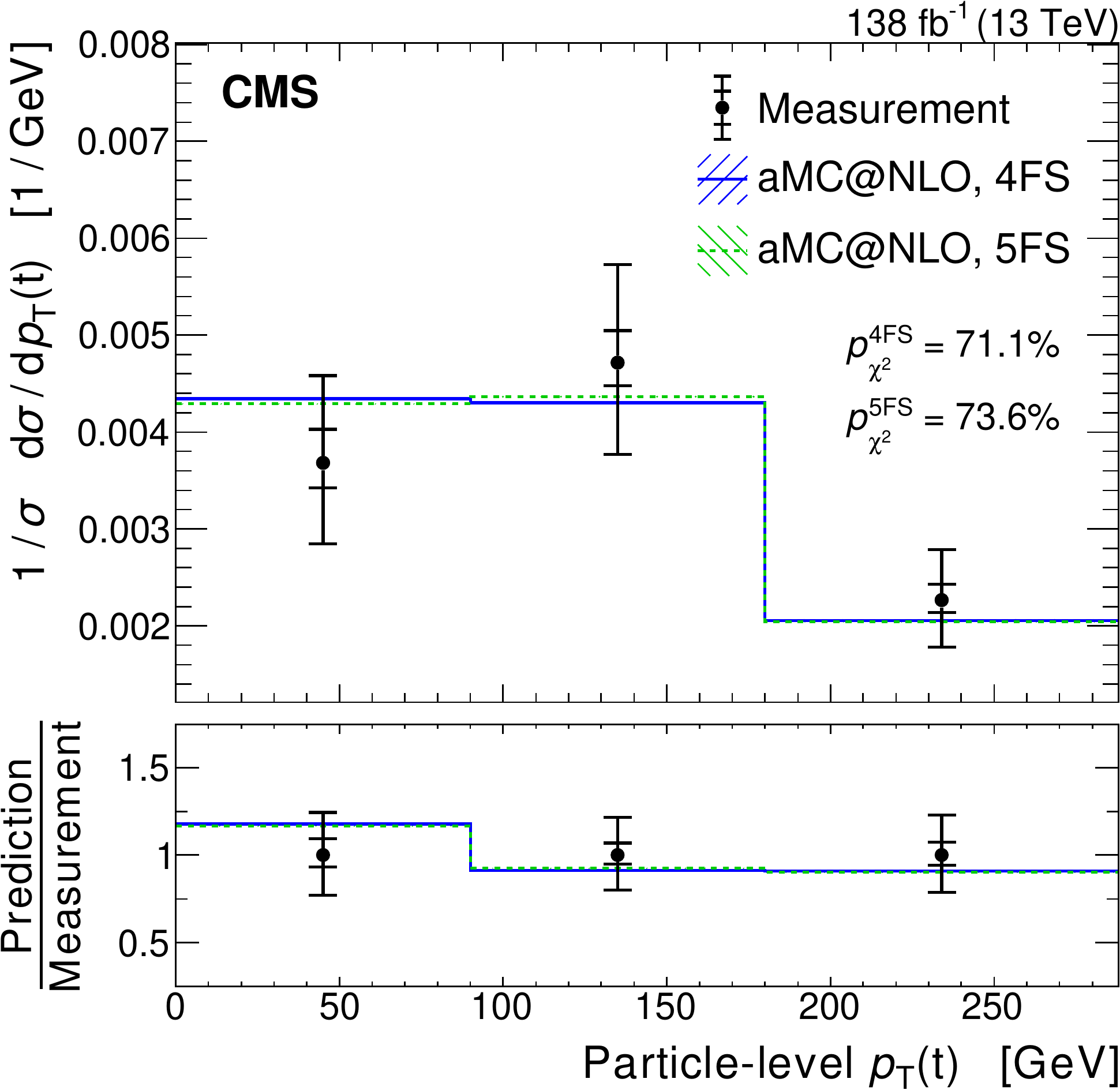}%
\captionof{figure}{%
    (Left) Absolute differential tZq production cross section as a function of the invariant mass of the three leptons~\cite{CMS-tZq}.
    (Right) Normalized differential tZq production cross section as a function of the top quark $p_{\mathrm{T}}$~\cite{CMS-tZq}.
}
\label{fig:tZq}
\end{center}

\section{Summary}

Cross section measurements of top quark production in association with a photon or a Z~boson, performed on proton-proton collision data recorded at \sqrts with the ATLAS and CMS experiments at the CERN LHC, are presented.
Very precise results are obtained for the inclusive production cross sections, and found in agreement with the expectation from the standard model of particle physics.
Various differential distributions are measured as well, and unfolding procedures are applied to present differential cross sections at parton- and particle-level.

%


\begin{thebibliography}{99}
\newcommand{\doi}[2]{\href{http://doi.org/#1}{#2}}
\newcommand{\arxiv}[1]{\href{http://arxiv.org/abs/#1}{arXiv:#1}}
\newcommand{\pas}[1]{\href{http://cms-results.web.cern.ch/cms-results/public-results/preliminary-results/#1/index.html}{CMS-PAS-#1}}
\newcommand{\etal}{\textit{et al.}}
\newcommand{\titl}[1]{}
\newcommand{\jrnl}[1]{\textit{#1}\xspace}
\newcommand{\vlme}[1]{\textbf{#1}\xspace}
\setlength{\itemsep}{-0.5\parskip}

\bibitem{ATLAS-Experiment}
ATLAS Collaboration,
\titl{The ATLAS experiment at the CERN Large Hadron Collider}%
\doi{10.1088/1748-0221/3/08/S08003}{\jrnl{JINST} \vlme{3} (2008) S08003}.

\bibitem{CMS-Experiment}
CMS Collaboration,
\titl{The CMS experiment at the CERN LHC}%
\doi{10.1088/1748-0221/3/08/S08004}{\jrnl{JINST} \vlme{3} (2008) S08004}.

\bibitem{ATLAS-ttG}
ATLAS Collaboration,
\titl{Measurements of inclusive and differential cross-sections of combined \ttbar\mygamma and tW\mygamma production in the e\mymu channel at 13\TeV with the ATLAS detector}%
\doi{10.1007/JHEP09(2020)049}{\jrnl{JHEP} \vlme{09} (2020) 049}
[\arxiv{2007.06946}].

\bibitem{theory-ttG1}
G.\ Bevilacqua \etal,
\titl{Hard photons in hadroproduction of top quarks with realistic final states}%
\doi{10.1007/JHEP10(2018)158}{\jrnl{JHEP} \vlme{10} (2018) 158}
[\arxiv{1803.09916}].

\bibitem{theory-ttG2}
G.\ Bevilacqua \etal,
\titl{Precise predictions for the $\text{\ttbar\mygamma}/\text{\ttbar}$ cross section ratios at the LHC}%
\doi{10.1007/JHEP01(2019)188}{\jrnl{JHEP} \vlme{01} (2019) 188}
[\arxiv{1809.08562}].

\bibitem{CMS-ttGsl}
CMS Collaboration,
\titl{Measurement of the inclusive and differential \ttbar\mygamma cross sections in the single-lepton channel and EFT interpretation at \sqrts}%
\doi{10.1007/JHEP12(2021)180}{\jrnl{JHEP} \vlme{12} (2021) 180}
[\arxiv{2107.01508}].

\bibitem{CMS-ttGdi}
CMS Collaboration,
\titl{Measurement of the inclusive and differential \ttbar\mygamma cross section and EFT interpretation in the dilepton channel at \sqrts}%
\pas{TOP-21-004},
CERN 2021.

\bibitem{CMS-tGq}
CMS Collaboration,
\titl{Evidence for the associated production of a single top quark and a photon in proton-proton collisions at \sqrts}%
\doi{10.1103/PhysRevLett.121.221802}{\jrnl{Phys.\ Rev.\ Lett.} \vlme{121} (2018) 221802}
[\arxiv{1808.02913}].

\bibitem{CMS-ttZ}
CMS Collaboration,
\titl{Measurement of top quark pair production in association with a Z~boson in proton-proton collision at \sqrts}%
\doi{10.1007/JHEP03(2020)056}{\jrnl{JHEP} \vlme{03} (2020) 056}
[\arxiv{1907.11270}].

\bibitem{ATLAS-ttZ}
ATLAS Collaboration,
\titl{Measurements of the inclusive and differential production cross sections of a top-quark--antiquark pair in association with a Z~boson at \sqrts with the ATLAS detector}%
\doi{10.1140/epjc/s10052-021-09439-4}{\jrnl{Eur.\ Phys.\ J.\ C} \vlme{81} (2021) 737}
[\arxiv{2103.12603}].

\bibitem{theory-ttZ}
A.\ Kulesza \etal,
\titl{Associated top quark pair production with a heavy boson: differential cross sections at NLO+NNLL accuracy}%
\doi{10.1140/epjc/s10052-020-7987-6}{\jrnl{Eur.\ Phys.\ J.\ C} \vlme{80} (2020) 428}
[\arxiv{2001.03031}].

\bibitem{ATLAS-tZq}
ATLAS Collaboration,
\titl{Observation of the associated production of a top quark and a Z~boson in pp collisions at \sqrts with the ATLAS detector}%
\doi{10.1007/JHEP07(2020)124}{\jrnl{JHEP} \vlme{07} (2020) 124}
[\arxiv{2002.07546}].

\bibitem{CMS-tZq}
CMS Collaboration,
\titl{Inclusive and differential cross section measurements of single top quark production in association with a Z~boson in proton-proton collisions at \sqrts}%
\arxiv{2111.02860} (submitted to \jrnl{JHEP}).

\bibitem{CMS-tZq-pred}
CMS Collaboration,
\titl{Measurement of the associated production of a single top quark and a Z~boson in pp collisions at \sqrts}%
\doi{10.1016/j.physletb.2018.02.025}{\jrnl{Phys.\ Lett.\ B} \vlme{779} (2018) 358}
[\arxiv{1712.02825}].

\bibitem{CMS-ttW}
CMS Collaboration,
\titl{Measurement of the cross section for top quark pair production in association with a W~or Z~boson in proton-proton collisions at \sqrts}%
\doi{10.1007/JHEP08(2018)011}{\jrnl{JHEP} \vlme{08} (2018) 011}
[\arxiv{1711.02547}].

\bibitem{ATLAS-ttW}
ATLAS Collaboration,
\titl{Measurement of the \ttbar Z and \ttbar W cross sections in proton-proton collisions at \sqrts with the ATLAS detector}%
\doi{10.1103/PhysRevD.99.072009}{\jrnl{Phys.\ Rev.\ D} \vlme{99} (2019) 072009}
[\arxiv{1901.03584}].

\bibitem{theory-ttW}
R.\ Frederix \& I.\ Tsinikos,
\titl{On improving NLO merging for \ttbar W production}%
\doi{10.1007/JHEP11(2021)029}{\jrnl{JHEP} \vlme{11} (2021) 029}
[\arxiv{2108.07826}].

\end{thebibliography}
\end{document}